# Pulsed Sagnac source of polarization-entangled photon pairs in telecommunication band

Heonoh Kim[1], Osung Kwon[2,†] & Han Seb Moon[1,*]

[1]Department of Physics, Pusan National University, Geumjeong-Gu, Busan 46241, South Korea

[2]National Security Research Institute, Daejeon 34044, South Korea

[†]E-mail: oskwon@nsr.re.kr

[*]Corresponding author: hsmoon@pusan.ac.kr

**Abstract**

We report a source of polarization-entangled photon pairs in the 1550-nm telecommunication band, which is based on non-collinear spontaneous parametric down-conversion in a periodically poled lithium niobate crystal pumped by picosecond pulses. This source is realized utilizing a polarization-based Sagnac interferometer employing a type-0 non-collinear quasi-phase-matching configuration. Polarization entanglement is verified through measurement of the polarization-correlation interference fringes with visibility >96% and by testing the experimental violation of the Clauser-Horne-Shimony-Holt (CHSH) form of Bell's inequality. The CHSH-Bell parameter $S$ is found to be $2.72 \pm 0.04$, with 18 standard deviations from the statistical uncertainty.

**Introduction**

Entangled photon-pair sources are widely recognized as key components for implementation of entanglement-based quantum information technology, such as linear optical quantum computing[1,2], quantum communication[3,4], and quantum metrology[5,6]. Recently, a need has arisen for efficient sources of entangled photons with high brightness and phase stability, for use in real-world applications of quantum technology. Over the past two decades, various source systems for entangled photon generation have been reported, which have been designed to satisfy the requirements of such quantum technology applications. These systems are based on spontaneous parametric down-conversion (SPDC)[7], optical fibres[8], semiconductors[9], silicon waveguides[10], and atomic ensembles[11]. Among them, type-II collinear SPDC employing quasi-phase matching (QPM) in a periodically poled $KTiOPO_4$ (PPKTP) crystal is the most widely adopted method for entangled photon generation, having the advantages of high brightness and robust phase stability[12-21]. In particular, entanglement generation through a polarization-based Sagnac interferometer is a highly favored method because of the intrinsic phase stability of the resultant entangled photons[16-22]. In that case, two photons from the SPDC must be orthogonally polarized to allow subsequent separation of two photons at the two interferometer outputs, so as to generate lossless polarization entanglement.

Polarization-entangled photons have been generated in the telecommunication band by utilizing both PPKTP crystals[20,21] and periodically poled lithium niobate (PPLN) waveguides[23-26] operated in pulsed and continuous modes. However, in many experimental implementations of telecommunication-band entangled photons, various problems that affect the detection device have arisen, such as low efficiency, high dark counts, and long dead time for single-photon detectors. In particular, the dead time for suppression of the after-pulse contributions in gated-mode operations severely limits the entangled photon-pair coincidence counting rate. Use of QPM-SPDC in a type-II PPLN waveguide is one of the preferred approaches towards generation of polarization-entangled photons in the telecommunication band[25,26]. However, the conversion efficiency is very low compared with that of type-0 QPM-SPDC, although the spectral bandwidth of the former is much narrower[27,28]. Therefore, combination of non-collinear QPM-SPDC in a type-0 PPLN with a highly stable interferometer is a promising method for efficient generation of polarization-entangled photon pairs with high brightness and intrinsic phase stability in the telecommunication band. Moreover, a polarization-entangled photon-pair source based on type-0 QPM-SPDC can span a relatively broad range of optical telecommunication bands, such as the S-, C-, and L-bands. Therefore, such a source can also be utilized in the full range of the 8-channel coarse wavelength-division multiplexing (CWDM) optical band. Recently, a type-0 non-collinear QPM-SPDC in a PPKTP crystal operated in continuous mode was used to generate highly bright polarization-entangled photons at 810-nm wavelength[29].

In this paper, we experimentally demonstrate a polarization-entangled photon-pair source in the 1550-nm telecommunication band, which is generated in a polarization-based Sagnac interferometer involving spatially well-separated photon pairs emitted from a type-0 PPLN crystal satisfying the non-collinear QPM-SPDC condition. To verify the polarization entanglement quality of the generated photon pairs, we measure the polarization-correlation interference fringes and demonstrate experimental violation of the Clauser-Horne-Shimony-Holt (CHSH) form of Bell's inequality[30]. We note that non-collinear geometry employing type-0 QPM-SPDC can significantly enhance the brightness of the photon-pair source, and also aid development of new methods applicable to experimental quantum information science and technology.

**Results**

In the experiment, we used a mode-locked picosecond fibre laser as the QPM-SPDC pumping source, which had a 3.5-ps pulse duration at a 775-nm centre wavelength with a 20-MHz repetition rate. Correlated photon pairs in the telecommunication band centred at 1550 nm were generated through a non-collinear QPM-SPDC process in a 10-mm-long type-0 PPLN crystal with a 19.2-μm grating period [Fig. 1(a)]. Two degenerate photons with identical polarization were emitted with a full-opening angle of 4.6° in the non-collinear regime, when the PPLN-temperature was set to 40°C (see Supplementary Figure S1 for details). The two correlated photons with broad spectral bandwidths were coupled to single-mode fibres (SMFs) and then measured by two detectors after passing through CWDM filters. The detectors were InGaAs/InP single-photon avalanche photodiodes (ID 210, ID Quantique) operated in gated mode. The detector dead time and detection efficiency were set to 5 μs and 15%, respectively.

The single and coincidence counting rates measured under these conditions are shown as functions of the average pump power in Figs. 1(b) and 1(c), respectively. From the singles and coincidence counting rates measured under a low average pump power of 1 mW, the pair generation probability per pulse was estimated to be 0.046 at 20-MHz pump-pulse repetition rate[18]. For continuous-wave-pumped SPDC photon-pair sources, the spectral brightness (the number of detected pairs per second per mW of pump beam in a given spectral bandwidth, Hz/mW/nm) is commonly used to evaluate the source performance. On the other hand, for pulse-pumped photon-pair sources, estimating the pair generation probability per pulse is more useful and practical method to evaluate the source performance (see Supplementary Table 1 for details).

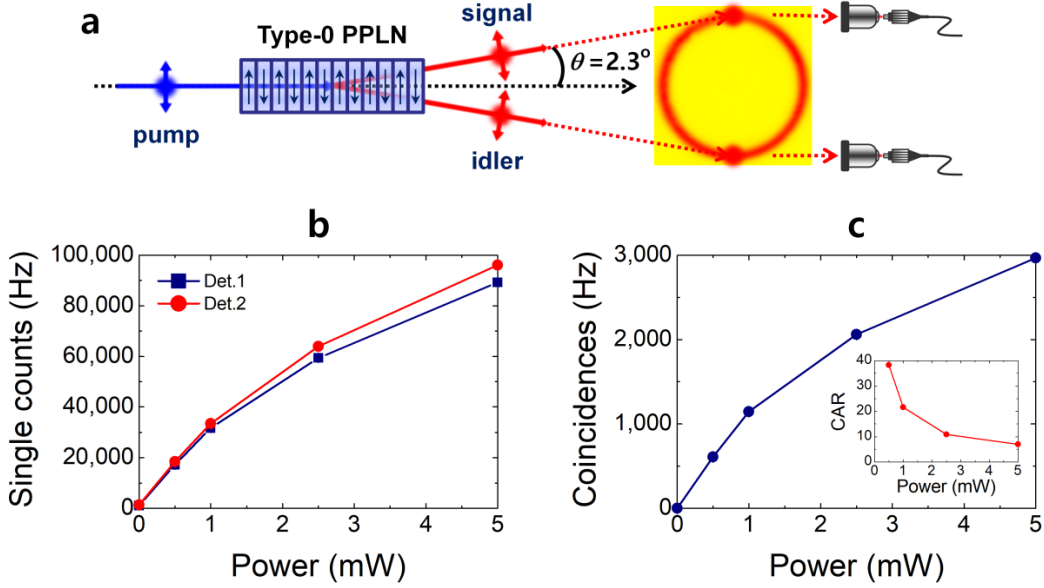

**Figure 1.** Generation of correlated photon pairs in PPLN crystal. (**a**) Correlated photon pairs in the telecommunication band are generated via non-collinear QPM-SPDC in a 10-mm-long type-0 PPLN crystal. The two photons are spatially well separated from the pump beam direction with a full-opening angle of 4.6°, and then coupled by single-mode fibres through the coupling optics. (**b**) Single and (**c**) coincidence counting rates as functions of pump power. Inset in **c** shows the coincidence-to-accidental-ratio (CAR) as a function of the pump power.

Furthermore, for a gated-mode operation of the single-photon counting detector, the maximum trigger rate is determined by the single counting rate and dead time of the detector, which is given by the relation, $f_{\text{trig. max.}} = f_{\text{trig. in}} - R_s \left( T_{\text{dead}} / T_{\text{trig.}} \right)$, where $f_{\text{trig. max.}}$ is the maximum trigger rate after detecting the input photons, $f_{\text{trig. in}}$ is the applied external trigger rate, $R_s$ is the measured single counting rate, $T_{\text{dead}}$ is a dead time of the detector, and $T_{\text{trig.}}$ is a period of the applied external trigger signal. In addition, the maximally obtainable single counting rate $R_{s, \text{max.}}$ is also limited by the ratio of $T_{\text{dead}}$ to $T_{\text{trig.}}$, $R_{s, \text{max.}} = f_{\text{trig. in}} / \left( T_{\text{dead}} / T_{\text{trig.}} \right)$. In our experimental conditions, the dead time of 5 μs and the external trigger rate of 20 MHz, the average single counting rate at the two detectors was measured to be approximately 32 kHz for 1 mW pump power. In this case, the maximum trigger rate becomes 16.8 MHz.

The SMF-coupled bandwidth of the generated photons was confirmed by measuring the Hong-Ou-Mandel interference fringe without spectral filters, and it was found to be approximately 132 nm (see Supplementary Figure S2 for details). In that case, the spectral purity was estimated to be 0.022 because of the rather narrow pump beam bandwidth and strong spectral correlation between the generated photons[31]. When a CWDM filter (18-nm bandwidth) was used in our experiment, the purity increased to 0.157. A very high spectral purity of 0.994 could be achieved with a tight spectral filtering of 1 nm (see Supplementary Figure S3 for details); this is comparable to the case in which type-II QPM is employed as the photon pair source[27].

**Experimental setup**. Figure 2 shows the experimental setup used in this study to generate polarization-entangled photons in a polarization-based Sagnac interferometer (see Supplementary Figure S4). The pump pulse was guided through an SMF and then collimated by a fibre coupler (FCp, where "p" indicates "pump"). Then, the pump pulse propagated into free-space with a full width at half maximum (FWHM) of approximately 1 mm. The pump beam was subsequently focused into the centre of the PPLN crystal through a spherical lens with 200-mm focal length, yielding an FWHM spot size of 60.5 μm and a Rayleigh length of 21.4 mm. A linear polarizer (Pp) and half-wave plate (Hp) were used to balance the average pump power at the two output ports of the dual-wavelength polarizing beam splitter (DPBS). The average pump power was set at 1 mW in front of the crystal.

The working principles of the polarization-based Sagnac interferometer scheme for polarization-entangled photon pair generation are as follows: (i) A pump beam with 45° polarization at the DPBS input port is split in two, with horizontal (*H*; clockwise, CW) and vertical (*V*; counterclockwise, CCW) polarization components: $|45^o\rangle_p \xrightarrow{\text{DPBS}} 1/\sqrt{2}(|H\rangle_p + |V\rangle_p)$. (ii) Horizontal polarization of the transmitted CW pump beam is transformed into vertical polarization via a dual-wavelength half-wave plate (DHWP) fixed at 45°, so as to generate vertically polarized photon pairs in the PPLN crystal: $|H\rangle_p \xrightarrow{\text{DHWP}} |V\rangle_p \xrightarrow{\text{PPLN}} |V\rangle_1|V\rangle_2$. (iii) Vertical polarization of the reflected CCW pump beam generates vertically polarized photon pairs in the same PPLN crystal, and the polarization direction is transformed to horizontal via the DHWP:

$|V\rangle_p \xrightarrow{PPLN} |V\rangle_1|V\rangle_2 \xrightarrow{DHWP} |H\rangle_1|H\rangle_2$. As a result, the two two-photon polarization states, $|V\rangle_1|V\rangle_2$ and $|H\rangle_1|H\rangle_2$, generated through the CW and CCW Sagnac loops, respectively, produce a polarization-entangled state.

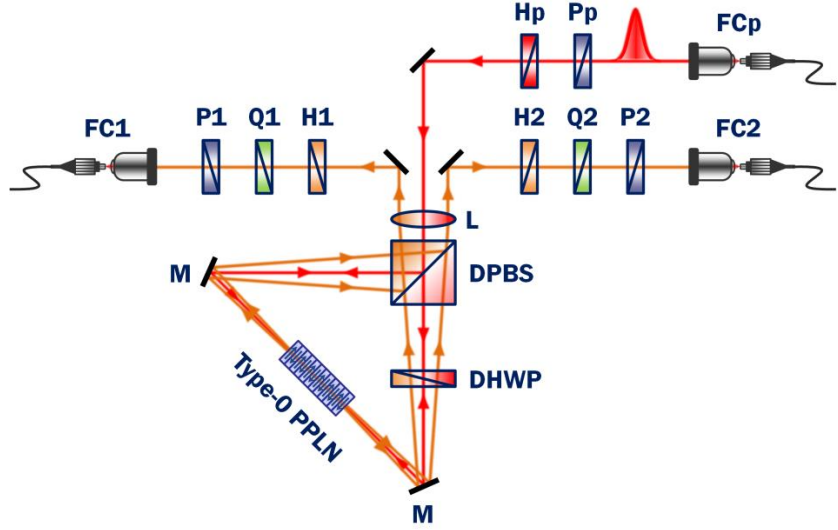

**Figure 2.** Experimental setup for generation of polarization-entangled photons in polarization-based Sagnac interferometer. Non-collinear photon pairs were generated in a type-0 PPLN crystal. Pump, mode-locked picosecond pulse laser (775 nm, 20 MHz, 3.5 ps); FC, SMF coupler for pump (p) and correlated photons 1 and 2; P, linear polarizer; H, half-wave plate; Q, quarter-wave plate; L, spherical lens (200-mm focal length); DPBS, dual-wavelength polarizing beam splitter; DHWP, dual-wavelength half-wave plate; M, mirror; PPLN, type-0 periodically-poled lithium niobate crystal with 10-mm length.

The two-photon polarization state at the Sagnac interferometer output ports is expressed in the form

$$|\Phi\rangle = \frac{1}{\sqrt{2}}\left(|H\rangle_1|H\rangle_2 + e^{i\phi}|V\rangle_1|V\rangle_2\right), \qquad (1)$$

where the subscripts denote the two spatial modes. The phase factor $\phi$ arises from the relative phase difference between the two two-photon amplitudes. In fact, this phase shift is mainly caused by spatial overlap of the two non-collinear emission spectra coupled into FC1 and FC2 (Fig. 2), which strongly depend on both the position of the PPLN crystal and the alignment of the CW and CCW optical paths. The phase offset can be partially compensated by tilting one of the wave plates relative to its transmission axis in the optical path.

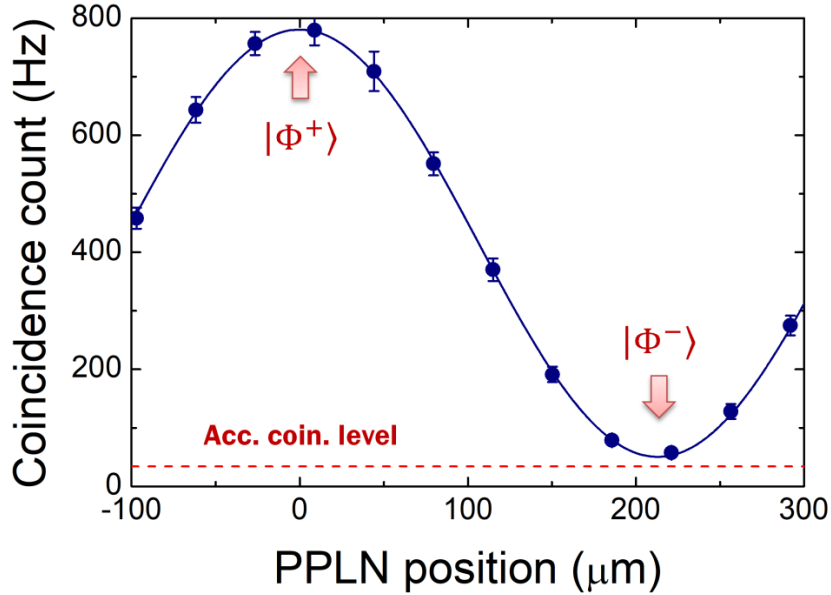

**Figure 3.** Compensation of phase offset by positioning of PPLN crystal. The initial condition of $\phi = 0$ was confirmed by the polarization-correlation measurements for the diagonal basis. When the coincidence counting rate was maximized the $|\Phi^+\rangle$ polarization-entangled state was prepared. Acc. coin. level: Accidental coincidence level.

For the non-collinear Sagnac source in our experiment, a total phase shift accumulated during focusing of the collimated pump beams, re-collimation of the down-converted photons via a lens, and FC coupling. This phase offset was caused by longitudinal walk-off and lack of spatial overlap between photon pairs generated through the CW and CCW Sagnac loops. The initial polarization-entangled state, $|\Phi^+\rangle$, was prepared by carefully adjusting the relative phase $e^{i\phi}$ between the two two-photon amplitudes ($|H\rangle_1|H\rangle_2$ and $|V\rangle_1|V\rangle_2$) in Eq. (1). The initial condition of $\phi = 0$ could be confirmed by measuring the polarization correlation for the diagonal basis. This is because the coincidence counting rate was maximized when the angles of the two polarizers (P1 and P2) were both set to +45°. Figure 3 shows the measured coincidence count as a function of the crystal position. The relative phase varied continuously depending on the crystal position; therefore, the initial polarization-entangled state changed from $|\Phi^+\rangle$ to $|\Phi^-\rangle$ accordingly. The filled circles represent experimental data and the solid line is a sinusoidal curve fit to the data points. The error bars denote the standard deviations of the measured raw coincidence events. The horizontal dashed line indicates the accidental coincidence counting level, which yields observed visibilities of 0.88 ± 0.01 and 0.97 ± 0.01 for the raw and net coincidences, respectively.

**Generation of polarization-entangled states.** In this experiment, to obtain the polarization-entangled state $|\Phi^+\rangle = 1/\sqrt{2}(|H\rangle_1|H\rangle_2 + |V\rangle_1|V\rangle_2)$ from Eq. (1), we adjusted the two optical paths toward the two FCs from the PPLN crystal by carefully tilting the two mirrors simultaneously. We also implemented compensation so that the phase offset was $\phi = 0$. Another polarization-entangled state, $|\Psi^+\rangle = 1/\sqrt{2}(|H\rangle_1|V\rangle_2 + |V\rangle_1|H\rangle_2)$, could also be easily generated by rotating one of the HWPs (H2 in Fig. 2) by 45°. The QWP (Q2) was used to adjust the phase factor to $\phi = \pi$ by rotating the angle by 90°, so as to generate $|\Phi^-\rangle$ and $|\Psi^-\rangle$ states from $|\Phi^+\rangle$. Polarization correlation measurements were performed as projective measurements of the coincidences at the FC1 and FC2 outputs, with relative-angle combinations of P1 and P2.

To verify the polarization entanglement quality of the photon pairs generated in our polarization-based Sagnac interferometer, we measured the polarization correlations in both the H/V and ±45° bases. Figure 4 shows the experimental results for the $|\Phi^+\rangle$, $|\Phi^-\rangle$, $|\Psi^+\rangle$ and $|\Psi^-\rangle$ states. For these polarization-entangled states, the coincidence counting probabilities were found to be $P(\Phi^\pm) = 1/2\cos^2(\theta_1 \mp \theta_2)$ and $P(\Psi^\pm) = 1/2\sin^2(\theta_1 \pm \theta_2)$, where $\theta_1$ and $\theta_2$ denote the rotation angles of P1 and P2, respectively, depending on the relative angle difference between the two polarizers. The single and coincidence counts between the two detectors were measured as functions of the P2 rotation angle with P1 fixed at both 0° (filled squares) and +45° (filled circles) for the H/V and ±45° correlations, respectively. In all cases, the measured single counting rates at the two individual detectors D1 and D2 were approximately 29,000 and 25,000

Hz, respectively. Accidental coincidences from unwanted multiple-pair generation events, $N_1 N_2 / f_{\text{trig.}}$, where $N_i$ is the single counting rate and $f_{\text{trig.}}$ is the trigger-signal repetition frequency, were subtracted from the recorded coincidences.

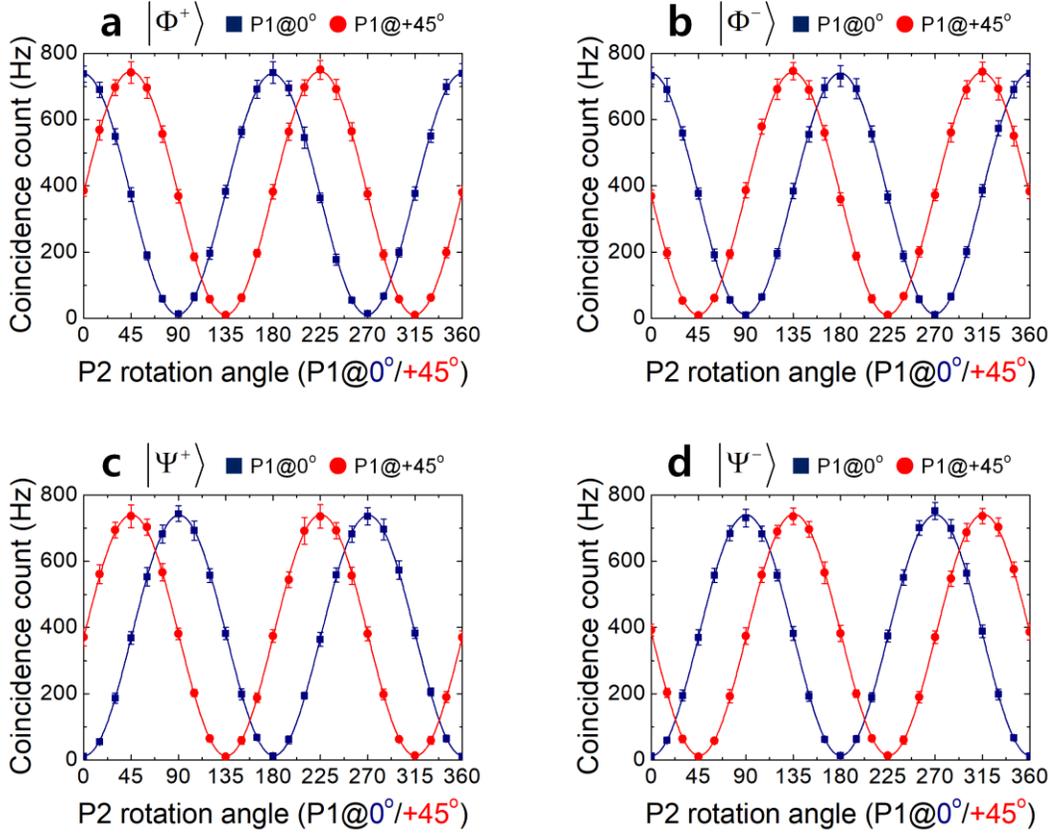

**Figure 4.** Experimental results. Polarization-correlation measurements for (**a,b**) $|\Phi^{\pm}\rangle$ and (**c,d**) $|\Psi^{\pm}\rangle$ states. The coincidence counts were measured as functions of the P2 rotation angle for P1 fixed at 0° and +45°. Accidental coincidences were subtracted from the measured coincidences. The solid lines show the theoretical fits.

For $|\Phi^+\rangle$, the observed polarization-correlation fringe visibilities were 0.97 ± 0.01 for both the *H/V* and ±45° bases, as shown in Fig. 4(a). In this figure, the symbols represent the net coincidence counts measured as functions of the P2 rotation angle, for P1 fixed at 0° and +45°. The solid lines show the sinusoidal fits. Without subtraction of the accidentals, the raw visibilities were 0.89 ± 0.01 for both the *H/V* and ±45° bases. The $|\Psi^+\rangle$ polarization-entangled state was obtained by rotating H2 by 45°. The same correlation measurements were performed to observe the polarization-correlation fringes, as shown in Fig. 4(c). For the ±45° bases, the observed visibilities were 0.88 ± 0.01 and 0.96 ± 0.01 for the raw and net coincidences, respectively. Similar visibilities were also obtained for the $|\Phi^-\rangle$ and $|\Psi^-\rangle$ states, as shown in Fig. 4(b) and 4(d), respectively. The observed high-visibility correlations, as shown for the ±45° basis, clearly illustrate the polarization entanglement of the photon-pair source. Further characterization of $|\Phi^+\rangle$ and $|\Psi^+\rangle$ states were performed via quantum state tomography[32]. The fidelities of the reconstructed density matrices with the states yield values of 0.851 (0.965) for $|\Phi^+\rangle$ and 0.834 (0.952) for $|\Psi^+\rangle$ states from the raw and net coincidences (see Supplementary Figure S5 for details).

**Experimental violation of CHSH-Bell inequality.** Next, we demonstrated the experimental violation of the CHSH-Bell inequality with the $|\Phi^+\rangle$ state. The CHSH-Bell inequality is defined as

$$S = \left| E(\theta_1, \theta_2) - E(\theta_1, \theta_2') + E(\theta_1', \theta_2) + E(\theta_1', \theta_2') \right| \leq 2, \qquad (2)$$

where $\theta_i$ are the rotation angles of the linear polarizers placed in front of the two FCs. Further, $E(\theta_1, \theta_2)$ is a polarization-correlation coefficient, which is expressed as

$$E(\theta_1, \theta_2) = \frac{C(\theta_1, \theta_2) - C(\theta_1, \overline{\theta_2}) - C(\overline{\theta_1}, \theta_2) + C(\overline{\theta_2}, \overline{\theta_2})}{C(\theta_1, \theta_2) + C(\theta_1, \overline{\theta_2}) + C(\overline{\theta_1}, \theta_2) + C(\overline{\theta_2}, \overline{\theta_2})}, \qquad (3)$$

where $C(\theta_1, \theta_2)$ is the coincidence counting rate between the two detectors when the two linear polarizers are set to $\theta_1$ and $\theta_2$, and $C(\theta_1, \overline{\theta_2})$ and $C(\overline{\theta_1}, \theta_2)$ represent the coincidence counting rates when one of the polarizers is rotated by an additional 90° relative to the rotation angles of $\theta_2$ and $\theta_1$, respectively. For the $|\Phi^+\rangle$ state, the coincidence counting rate in Eq. (3) is given by $C(\theta_1, \theta_2) \propto 1/2 \cos^2(\theta_1 - \theta_2)$ as shown in Fig. 4. Eq. (3) can also be expressed in a simpler form using the coincidence counting rates, $E(\theta_1,\theta_2) = \cos[2(\theta_1-\theta_2)]$. For example, if the relative angle between the two polarizers is set to 22.5°, the correlation coefficient in Eq. (3) is $1/\sqrt{2}$. In that case, we can obtain the maximum value for the CHSH-Bell parameter $S$ in order to violate the CHSH-Bell inequality. To estimate $S$, in this study, we measured the polarization-correlation coefficients for the $|\Phi^+\rangle$ state (see Supplementary Figure S6 for details). When the angles of the two polarizers were set to $\theta_1 = 0°$, $\theta_2 = 22.5°$, $\theta_1' = 45°$, and $\theta_2' = 67.5°$, the experimentally obtained correlation coefficients were found to be $E(\theta_1,\theta_2) = 0.6857 \pm 0.0229$, $E(\theta_1,\theta_2') = -0.6797 \pm 0.0179$, $E(\theta_1',\theta_2) = 0.6795 \pm 0.0148$, and $E(\theta_1',\theta_2') = 0.6745 \pm 0.0225$, which yields the CHSH-Bell parameter $S = 2.7194 \pm 0.0396$. Without subtraction of the accidental coincidences, the raw $S$ value was $2.4878 \pm 0.0370$. This result is in good agreement with the expected value $S = 2\sqrt{2}V_{coin.}$, where $V_{coin.}$ is the fringe visibility, as shown in Fig. 4. Thus, a violation of the CHSH-Bell inequality by 18.1667 standard deviations from the statistical uncertainty is clearly shown. If the measured coincidence counting rates exhibited Poissonian statistics, the standard deviation was found to be 19.0066. This value was obtained from the equation $\Delta S = \sqrt{(2-V_{coin.}^2)/N_{coin.}}$, where $N_{coin.}$ is the maximum coincidence rate measured in a given measurement basis.

**Conclusion**

In conclusion, we have experimentally demonstrated a pulsed polarization-entangled photon-pair source in the 1550 nm telecommunication band, which was generated in a polarization-based Sagnac interferometer. Spatially well-separated entangled photons were successfully extracted from a type-0 PPLN crystal satisfying the non-collinear QPM-SPDC condition. The entanglement quality was verified by measuring the polarization correlations, which indicated the fringe visibilities >88% and >96% for the raw and net coincidences, respectively. In addition, experimental violation of CHSH-Bell inequality was demonstrated, for which the parameter $S = 2.72 \pm 0.04$ was obtained, with 18 standard deviations from the statistical uncertainty. The proposed pulsed polarization-entangled photon-pair source can be used as an efficient and practical resource in fibre-based long-distance quantum communication applications and in many quantum optical experiments. Moreover, our scheme utilizing non-collinear geometry employing type-0 QPM-SPDC materials can aid generation of high-brightness polarization-entangled photons from photon pairs with identical polarizations.

**Acknowledgements**

This work was supported by the Basic Science Research Program through the National Research Foundation of Korea funded by the Ministry of Education, Science and Technology (No. 2018R1A2A1A19019181 and No. 2016R1D1A1B03936222).

**Author Contributions**

H.K., O.K. and H.S.M. conceived the project. H. K. designed the experimental setup and performed the experiment. H.K., O.K. and H.S.M. discussed the results and contributed to writing the manuscript. Correspondence and requests for materials should be addressed to O.K. (email: oskwon@nsr.re.kr) or H.S.M. (email: hsmoon@pusan.ac.kr)

**Additional Information**

**Competing Interests:** The authors declare no competing interests.

# Pulsed Sagnac source of polarization-entangled photon pairs in telecommunication band

Heonoh Kim,[1] Osung Kwon,[2,†] Han Seb Moon[1,*]

[1]Department of Physics, Pusan National University, Geumjeong-Gu, Busan 46241, South Korea
[2]National Security Research Institute, Daejeon 34044, South Korea
[†]E-mail: oskwon@nsr.re.kr
[*]Corresponding author: hsmoon@pusan.ac.kr

**Supplementary Figures and Table**

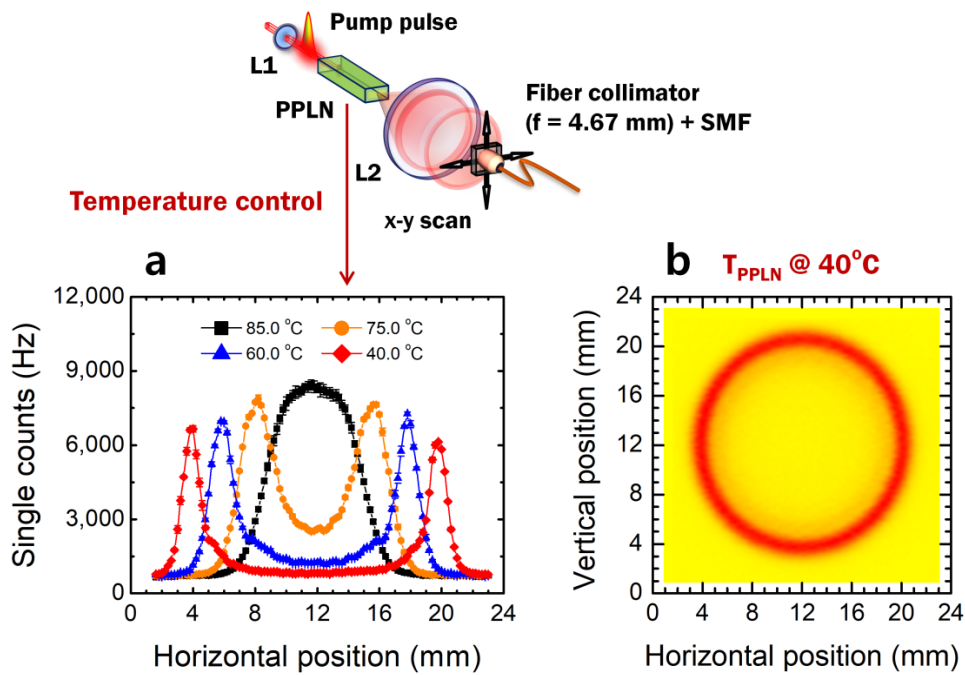

**Figure S1. Imaging of non-collinear QPM-SPDC spectra**. Correlated photon pairs in the 1550-nm telecommunication band are generated via non-collinear QPM-SPDC in a 10-mm-long type-0 PPLN crystal. L1 and L2 are spherical lenses with 200-mm focal length and diameters of 1 and 2 inch, respectively. Spatially distributed emission spectra are collected by imaging optics positioned after collimating lens L2. A bandpass filter with 10-nm bandwidth is attached in front of the imaging optics, which consists of SMF-collimator employing an aspherical lens with a 4.67-mm focal length (Numerical aperture: 0.53) and a two-dimensional translation stage. (**a**) Measured single-photon counting rates as functions of coupling-optics horizontal position for various PPLN temperatures. In this measurement, the vertical position of the collimator is fixed at the center position ($y = 12$ mm). (**b**) Non-collinear emission spectra measured with the imaging optics behind L2 when the PPLN temperature ($T_{PPLN}$) was set to 40°C. The imaging-optics step size was 0.2 mm in both the horizontal and vertical positions.

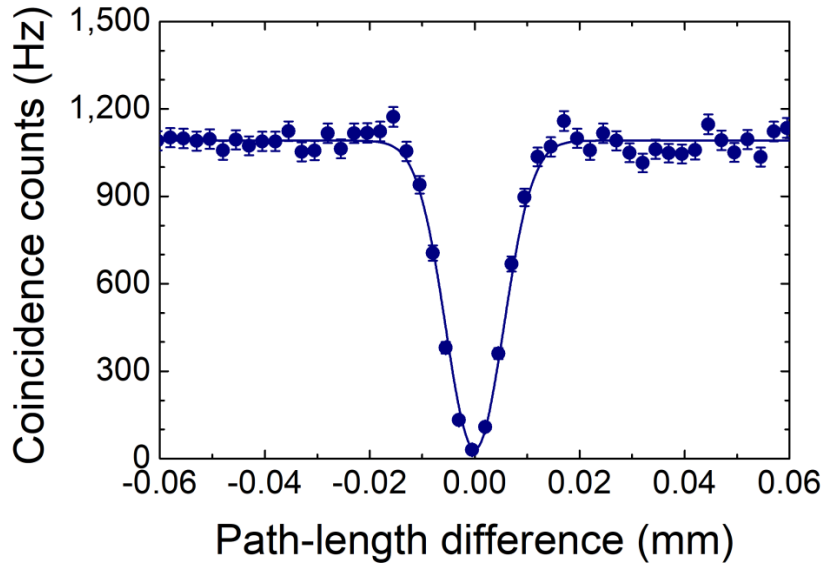

**Figure S2. Hong-Ou-Mandel interference fringe measured without filters.** Broadband spectra were emitted from the PPLN crystal under the condition of a non-collinear type-0 QPM SPDC. The SMF-coupled bandwidth of the generated photons was found to be approximately 132 nm, which was estimated from the Hong-Ou-Mandel interference-fringe width. The SMF coupling optics used in this measurement was also employed in the Sagnac source.

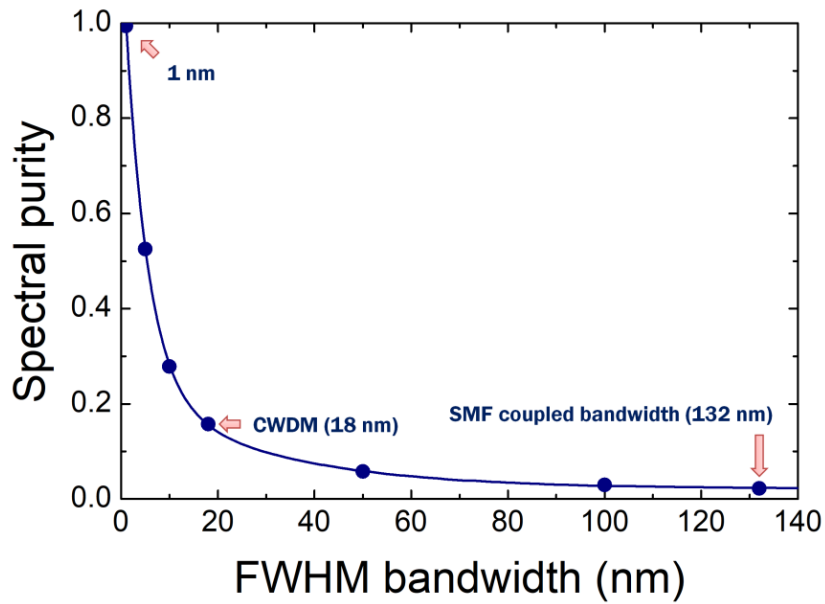

**Figure S3. Spectral purity vs spectral bandwidth.** The spectral purity of the SMF-coupled photon-pair source was estimated from the Schmidt numbers for given spectral bandwidths.

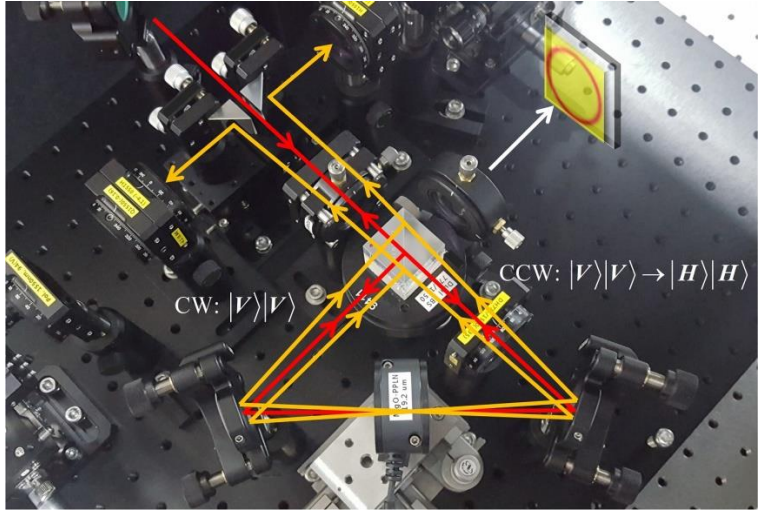

**Figure S4. Photograph of polarization-Sagnac source**. All optical components for source construction were positioned on 45 x 60 cm² breadboard. The Sagnac loop area was 10 x 10 cm².

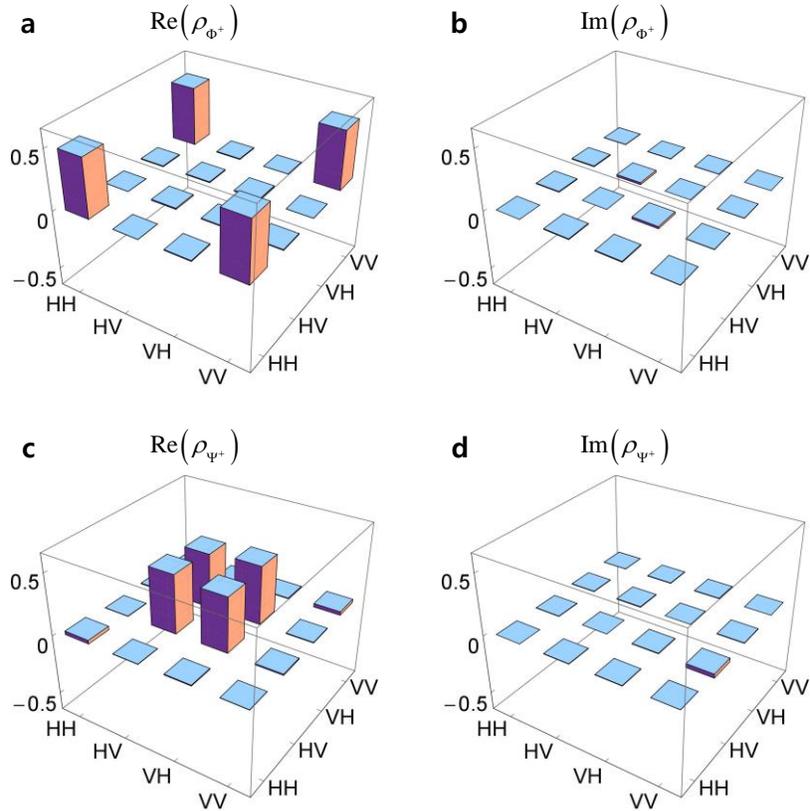

**Figure S5. Reconstructed density matrices for $|\Phi^+\rangle$ and $|\Psi^+\rangle$ states.** To verify the entanglement quality of the $|\Phi^+\rangle$ and $|\Psi^+\rangle$ states, we performed the quantum state tomography with coincidence events obtained from 16 measurement bases. (**a,c**) Real and (**b,d**) imaginary parts of the reconstructed density matrices. The fidelities of the reconstructed density matrices were 0.851 (0.965) for $|\Phi^+\rangle$ state and 0.834 (0.952) for $|\Psi^+\rangle$ state from the raw (net) coincidences, respectively.

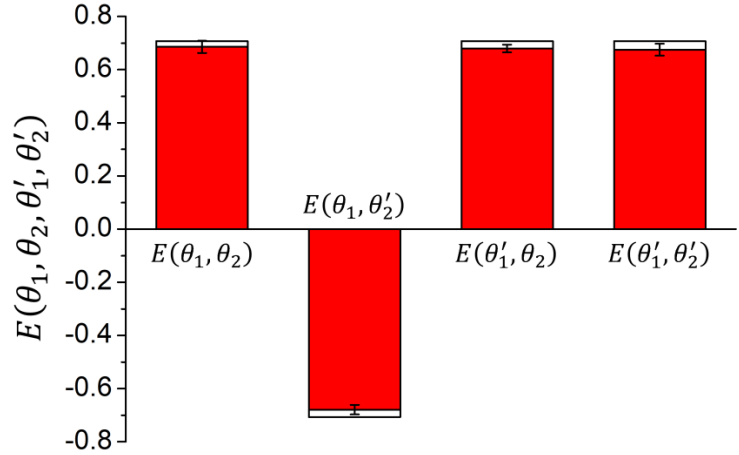

**Figure S6. Polarization-correlation coefficients for $|\Phi^+\rangle$ state.** To verify the experimental violation of the CHSH-Bell inequality for the $|\Phi^+\rangle$ state, we measured the correlation coefficients in Eq. (3) of the main text, when the angles of the two polarizers were set to $\theta_1 = 0°$, $\theta_2 = 22.5°$, $\theta_1' = 45°$, and $\theta_2' = 67.5°$. To obtain identical absolute heights for the histograms, the angle offsets of the two linear polarizers were carefully adjusted for the polarization-correlation measurements. The filled (blank) areas represent the measured (ideal) correlation coefficients and the error bars denote the standard deviations estimated from the measured coincidence counting uncertainties. The experimentally obtained correlation coefficients were $E(\theta_1, \theta_2) = 0.6857 \pm 0.0229$, $E(\theta_1, \theta_2') = -0.6797 \pm 0.0179$, $E(\theta_1', \theta_2) = 0.6795 \pm 0.0148$, and $E(\theta_1', \theta_2') = 0.6745 \pm 0.0225$, which yielded a CHSH-Bell parameter $S$ value of $2.7194 \pm 0.0396$. Without subtraction of the accidental coincidences, the raw $S$ value of the CHSH-Bell inequality was $2.4878 \pm 0.0370$.

**Table 1.** Comparison of the polarization-entangled photon-pair sources based on the quasi-phase matched (QPM) SPDC, which are implemented in the Sagnac configuration. [*]PGR, pair generation rate per pulse; [*]SNSPD, superconducting nanowire single-photon detector; [*]c-SHG/SPDC, cascaded SHG and the following SPDC.

| SPDC type | Phase matching Detector | Visibility/ Fidelity | Brightness (Hz/mW/nm) | References |
|---|---|---|---|---|
| Type-II Collinear QPM PPKTP, 10 mm | 405 nm (cw) → 810/810 nm Si-APD | V = 96.8% | 5 x $10^3$ | Kim et al.[1] |
| Type-II Collinear QPM PPKTP, 25 mm | 405 nm (cw) → 810/810 nm Si-APD | V = 99.5% | 2.73 x $10^5$ | Fedrizzi et al.[2] |
| Type-0 Collinear QPM PPLNWG, 1 mm | 776 nm (300 fs, 73 MHz) → 1542/1562 nm InGaAs-APD | F = 0.863 | [*]PGR = 0.185 | Lim et al.[3] |
| Type-0 Collinear QPM PPLNWG, 10 mm | 1548.6 nm (120 ps, 40 MHz) → 1538.8/1558.66 nm [*]c-SHG/SPDC InGaAs-APD | V = 96% F = 0.97 | [*]PGR = 0.04 | Arahira et al.[4] |
| Type-II Collinear QPM PPKTP, 15 mm | 404 nm (2 ps, 76 MHz) → 808/808 nm Si-APD | V = 98.7% F = 0.982 | 1.78 x $10^4$ | Predojević et al.[5] |
| Type-II Collinear QPM PPKTP, 30 mm | 792 nm (2 ps, 76 MHz) → 1584/1584 nm [*]SNSPD | V = 96% F = 0.98 | [*]PGR = 0.014 | Jin et al.[6] |
| Type-II Collinear QPM PPKTP, 10 mm | 775 nm (cw) → 1550/1550 nm InGaAs-APD | V = 96.4% F = 0.935 | 2 | Li et al.[7] |
| Type-0 Non-collinear QPM PPKTP, 30 mm | 405 nm (cw) → 810/810 nm Si-APD | V = 97% F = 0.975 | 0.025 | Jabir et al.[8] |
| Type-0 Non-collinear QPM PPLN, 10 mm | 775 nm (3.5 ps, 20 MHz) → 1550/1550 nm InGaAs-APD | V = 96% F = 0.965 | [*]PGR = 0.046 | Our work |